\documentclass{aa}
\usepackage{psfig,times}

\newcommand{\be}{\begin{equation}}
\newcommand{\ee}{\end{equation}}
\newcommand{\bea}{\begin{eqnarray}}
\newcommand{\eea}{\end{eqnarray}}

\def\apj{ApJ}
\def\apjs{ApJS}

\def\aea{A\&A}

\def\aas{A\&AS}

\def\la{\mathrel{\hbox{\rlap{\hbox{\lower4pt\hbox{$\sim$}}}\hbox{$<$}}}}
\def\ga{\mathrel{\hbox{\rlap{\hbox{\lower4pt\hbox{$\sim$}}}\hbox{$>$}}}}

\def\la{\mathrel{\hbox{\rlap{\hbox{\lower4pt\hbox{$\sim$}}}\hbox{$<$}}}}
\def\ga{\mathrel{\hbox{\rlap{\hbox{\lower4pt\hbox{$\sim$}}}\hbox{$>$}}}}

\begin{document}

\title {
High-resolution X-ray spectroscopy of Procyon by Chandra and XMM-Newton
}

\author{ 
        A.J.J. Raassen\inst{1,2}
        \and
        R. Mewe\inst{1}
        \and
        M. Audard\inst{3}
        \and
        M. G\"udel\inst{3}
        \and
        E. Behar\inst{4}
        \and
        J.S. Kaastra\inst{1}
        \and
        R.L.J. van der Meer\inst{1}
        \and
        C.R. Foley\inst{5}
        \and
        J.-U. Ness\inst{6}
}
\institute{
 SRON National Institute for Space Research,
 Sorbonnelaan 2, 3584 CA Utrecht, The Netherlands
  \and
 Astronomical Institute ``Anton Pannekoek'', Kruislaan 403,
 1098 SJ Amsterdam, The Netherlands 
  \and
 Paul Scherrer Institut, W\"urenlingen \& Villigen, 5232 Villigen PSI, Switzerland 
 \and
 Columbia Astrophysics Laboratory, Columbia University,
 New York, NY 10027, USA
  \and
 Mullard Space Science Laboratory, University College London,
 Surrey, RH5 6NT, United Kingdom 
 \and
 Universit\"at Hamburg, Gojenbergsweg 112, 21029 Hamburg, Germany
} 
\offprints {A.J.J.Raassen,\\
\email a.j.j.raassen@sron.nl}

\date{Received date \today; accepted date $\ldots$}

\abstract{
We report the analysis of the high-resolution soft X-ray spectrum of the nearby F-type star Procyon in the wavelength range from 5 to 175 \AA\ 
obtained with the Low Energy Transmission Grating Spectrometer (LETGS) on board Chandra and with 
the Reflection Grating Spectrometers (RGS) and the EPIC-MOS CCD spectrometers on board XMM-Newton. 
Line fluxes have been measured separately for the RGS and LETGS. Spectra have been fitted globally to
obtain self-consistent temperatures, emission measures, and abundances.
The total volume emission measure is $\sim 4.1 \times 10^{50}$ cm$^{-3}$ with a peak between 1 and 3 MK.
No indications for a dominant hot component ($T \ga$~4~MK)
were found. We present additional evidence for the lack of a solar-type FIP-effect, confirming earlier EUVE results.
}

\maketitle


\section{Introduction}
Magnetized hot outer atmospheres (coronae) are ubiquitous in
cool stars (spectral classes F-M). The particular example of the 
solar corona has revealed rich details on coronal structures, 
thermal stratification, abundance patterns, and the physics of
heating and mass motion. Nevertheless, in many other stars coronal
phenomena not common to the Sun are regularly observed, such as
persistent high-density ($n_\mathrm{e}>10^{10}$~cm$^{-3}$) coronal plasmas,
persistent very hot gas ($T>10$~MK), or abundances at 
variance with solar values (Bowyer et al. 2000). The Sun's relatively modest magnetic activity
is representative for a particular evolutionary state of a 1$M_{\odot}$
star (G\"udel et al. 1997), while most stellar objects regularly observed by 
X-ray satellites belong either to the more abundant low-mass classes
with some exceptional magnetic activity (M dwarfs and young K dwarfs) or
to tidally coupled binary systems with strongly enhanced magnetic
dynamos (RS CVn or Algol-type binaries). 

High-resolution X-ray spectroscopy
of such stellar systems now available from Chandra and XMM-Newton (Brinkman et al. 2000, 2001) 
has revealed coronal features clearly at variance with solar phenomena.
However, to translate solar knowledge to stellar environments, it is 
important to study stars that are relatively similar to the Sun. Given the
low X-ray luminosity of such stars, there are only a few in
the solar neighborhood accessible to high-resolution X-ray spectroscopy.
We present here a detailed analysis of the X-ray spectrum of Procyon, a nearby bright X-ray
source with a coronal plasma of about 1--3~MK, exhibiting a cooler X-ray spectrum than
magnetically active stars that have predominantly
been studied so far with XMM-Newton and Chandra (Audard et al. 2001ab; G\"udel et al. 2001ab; Mewe et al. 2001).
The late-type (F5 IV-V) optically bright (m$_\mathrm{v}$=0.34) star Procyon (with a faint white dwarf companion) at a 
distance of 3.5 pc has a line-rich coronal spectrum in the X-ray region. The mass of Procyon is 1.75$M_\odot$ and its 
radius 2.1$R_\odot$ (Irwin et al. 1992). The high-resolution spectrum of Procyon has been studied earlier using EUVE 
(Drake et al. 1995; Schrijver et al. 1995; Schmitt et al. 1996) and by means of the LETGS on board Chandra (Ness et al. 2001). 
Ness et al. focussed their efforts on the density-sensitive and temperature-sensitive lines of C V, N VI, and O VII only.
Here we present an extended investigation of the LETGS spectrum covering the total spectral range from 5-175 \AA\ together with 
the analysis of the RGS spectra from 5-37~\AA.
In Sect.~2 we describe the observations. Sect.~3 (Analysis) is divided into a part on global fitting (3.2) based on
the total spectrum and a part that contains consistency checks based on individual line measurements (3.3). 

\begin{figure*}
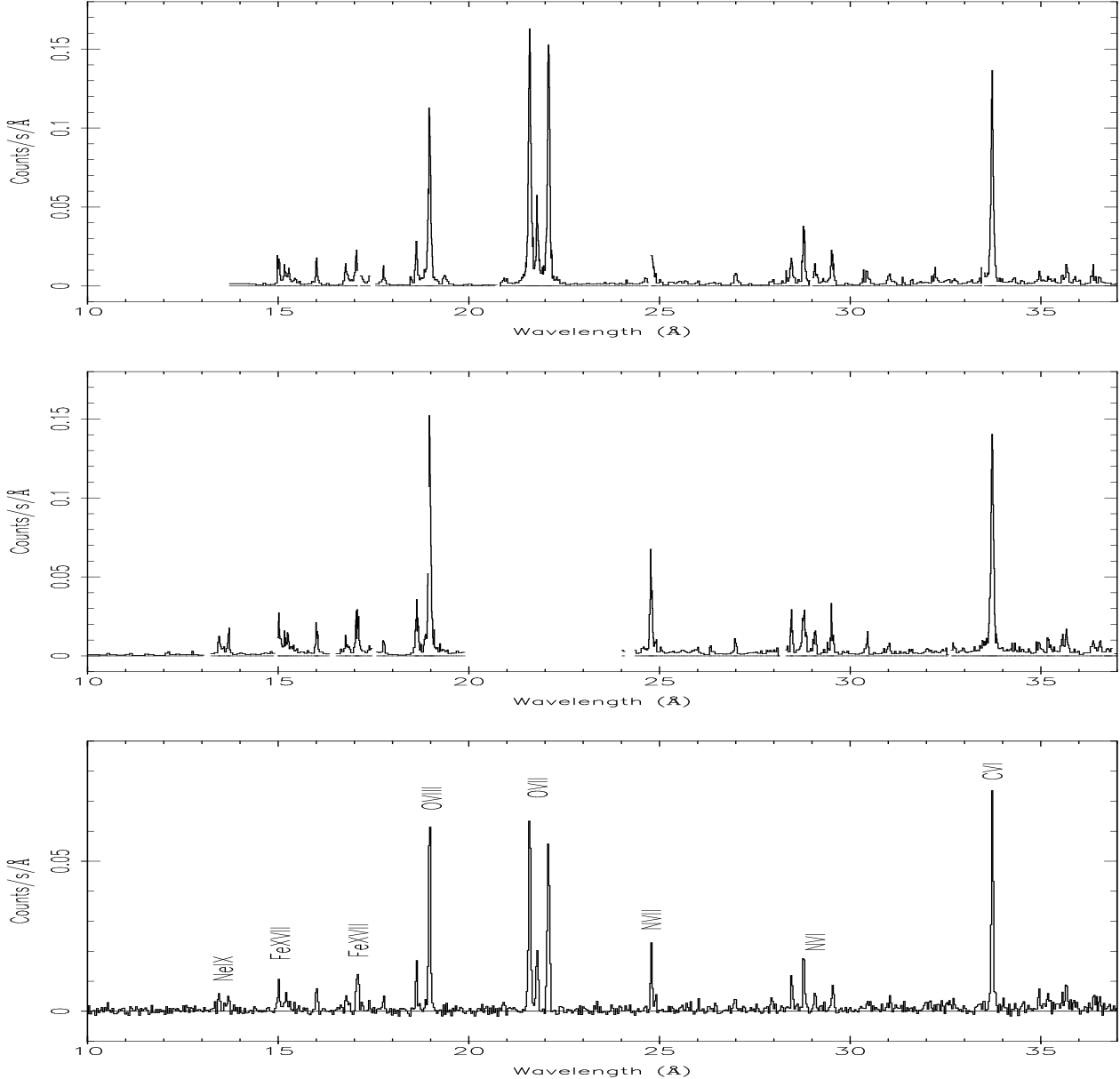

\vbox{
\hbox{
\psfig{figure=H3110F1a.ps,angle=-90,height=5.5truecm,width=16.7truecm,clip=}}
\hbox{
\psfig{figure=H3110F1b.ps,angle=-90,height=5.5truecm,width=16.7truecm,clip=}}
\hbox{
\psfig{figure=H3110F1c.ps,angle=-90,height=5.5truecm,width=16.7truecm,clip=}}
}
\caption[]{From top to bottom the spectra of Procyon observed by RGS1, RGS2 and LETGS 
in the wavelength region from 10 to 37 \AA
}
\end{figure*}

\begin{figure*}
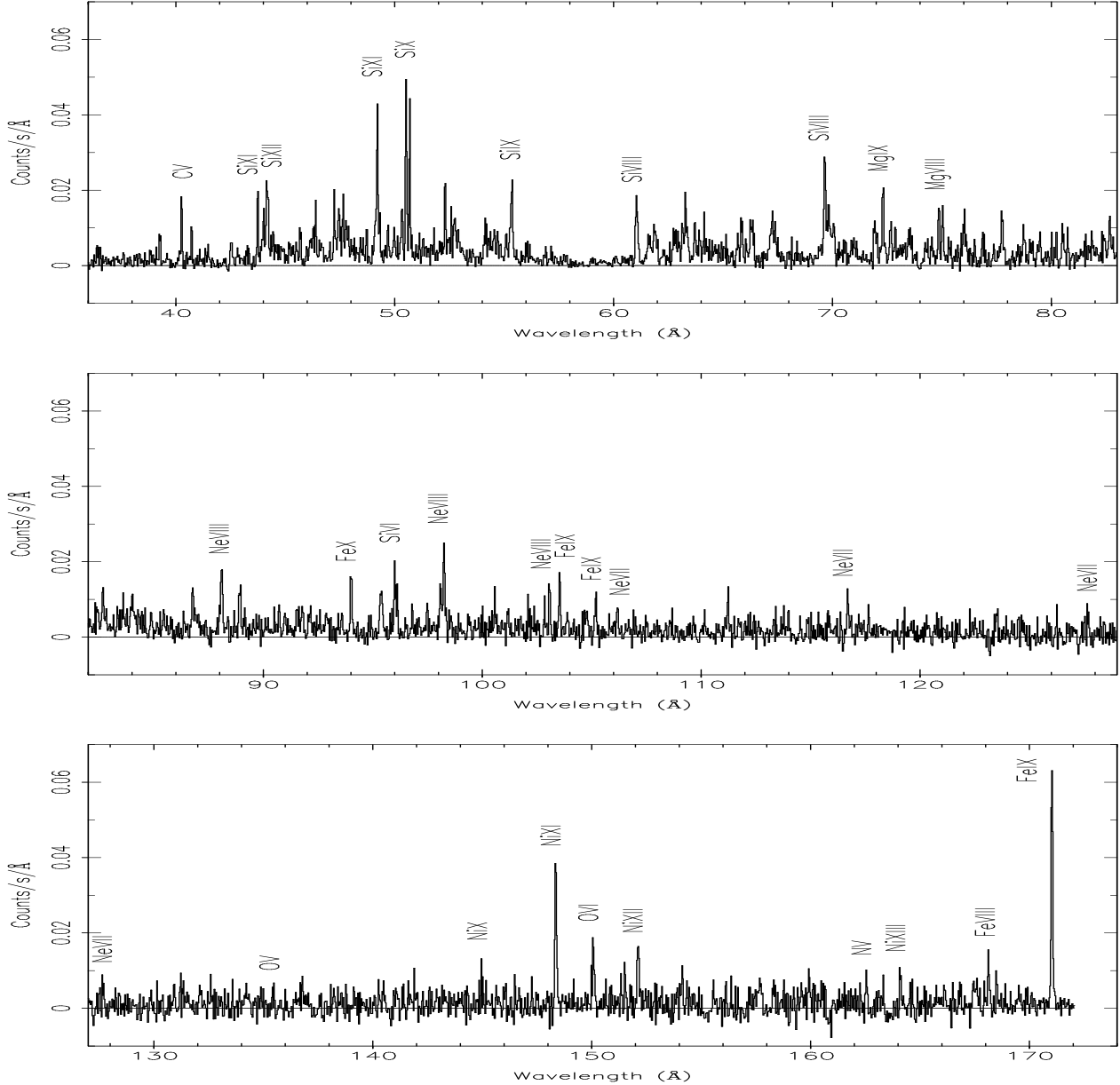

\center{\psfig{figure=H3110F2a.ps,angle=-90,height=5.53truecm,width=16.7truecm,clip=}}
\center{\psfig{figure=H3110F2b.ps,angle=-90,height=5.53truecm,width=16.7truecm,clip=}}
\center{\psfig{figure=H3110F2c.ps,angle=-90,height=5.53truecm,width=16.7truecm,clip=}}
\caption[]{The Procyon spectrum observed by LETGS 
in the region 36-174 \AA}

\end{figure*}

\begin{table*}
\caption{Wavelengths and fluxes for RGS1, RGS2, and LETGS, together with
the line identifications from MEKAL and KELLY}         
\begin {center}
\label{upper1}
\begin{tabular}{lllllllllr}
\hline 
\multicolumn{2}{c|}{RGS1}&\multicolumn{2}{c|}{RGS2}&\multicolumn{2}{c|}{LETGS}&\multicolumn{4}{c}{Line identifications$^a$}\\
\hline
$\lambda$(\AA) &  flux$^b$&  $\lambda$(\AA) & flux$^b$ &$\lambda$(\AA) & flux$^b$ &  MEKAL &flux$^c$ &  KELLY  &    Ion \\
\hline
   gap  & ---  &  13.454(9) & 0.17(4) &  13.450(13) & 0.14(5) &  13.448&0.14 &  13.447 &    Ne IX \\ 
   gap  & ---  &  13.701(6) & 0.17(4) &  13.690(15) & 0.12(5) &  13.700&0.10 &  13.700 &    Ne IX \\ 
 15.018(7) & 0.20(5) &  15.024(9) & 0.26(4) &  15.015(18) & 0.21(6) &  15.014&0.16 &  15.013 &  Fe XVII$^d$ \\ 
 15.176(13) & 0.12(3) &  15.161(21) & 0.08(3) &    ---  & ---  &  15.175&0.03 &  15.176 &   O VIII \\ 
  ---   & ---  &     --- &  --- &  15.207(28) & 0.13(5) &     ---&0.09 &     --- &  Fe XVII$^e$ \\ 
 15.281(12) & 0.11(3) &  15.258(12) & 0.09(5) &    ---  &  --- &  15.265&0.04&  15.260 &  Fe XVII \\ 
 16.008(3) & 0.17(3) &  16.013(9) & 0.20(5) &  16.008(9) & 0.18(5) &  16.003&0.15 &  16.007 &   O VIII$^f$ \\ 
 16.776(9) & 0.16(3) &  16.776(15) & 0.12(4) &  16.790(14) & 0.13(5) &  16.780&0.10 &  16.775 &  Fe XVII \\ 
 17.047(9) & 0.23(4) &  17.044(14) & 0.19(5) &  17.054(12) & 0.22(9) &  17.055&0.12 &  17.051 &  Fe XVII \\ 
   ---  & ---  &  17.099(12) & 0.25(5) &  17.102(9) & 0.29(9) &  17.100&0.10 &  17.100 &  Fe XVII \\ 
 17.402(12) & 0.07(3) &  17.402(28) & 0.06(3) &  17.396(17) & 0.08(5) &  17.380&0.04&  17.396 &    O VII \\ 
 17.765(9) & 0.10(3) &  17.772(9) & 0.11(4) &  17.769(13) & 0.14(5) &  17.770&0.12 &  17.768 &    O VII \\ 
 18.624(4) & 0.30(4) &  18.637(6) & 0.37(5) &  18.629(5) & 0.39(8) &  18.628&0.39 &  18.627 &    O VII \\ 
 18.970(2) & 1.61(8) &  18.975(2) & 1.78(11) &  18.972(2) & 1.83(15) &  18.973&1.93 &  18.969 &   O VIII \\ 
 20.918(27) & 0.04(3) &    gap  & ---  &  20.905(22) & 0.14(7) &  20.910&0.06 &  20.910 &    N VII \\ 
 21.596(2) & 2.36(11) &    gap  & ---  &  21.597(2) & 3.01(25) &  21.602&3.35 &  21.602 &    O VII \\ 
 21.797(4) & 0.53(6) &    gap  & ---  &  21.792(5) & 0.90(14) &  21.800&0.80 &  21.804 &    O VII \\ 
 22.098(2) & 2.20(11) &    gap  & ---  &  22.089(2) & 2.57(23) &  22.100&2.33 &  22.101 &    O VII \\ 
    ---    &  ---     &  24.780(3) & 0.88(7) &  24.790(4) & 0.80(14) &  24.781&0.76 &  24.781 &    N VII \\ 
    ---    &  ---     &  24.907(22) & 0.08(4) &  24.906(11) & 0.18(9) &  24.900&0.08 &  24.898 &     N VI \\ 
 27.001(11) & 0.16(4) &  26.994(12) & 0.14(4) &  26.979(19) & 0.18(9) &  26.990&0.08 &  26.990 &     C VI \\ 
 28.460(7) & 0.30(5) &  28.465(6) & 0.39(6) &  28.470(6) & 0.49(12) &  28.470&0.39 &  28.466 &     C VI \\ 
 28.785(4) & 0.69(9) &  28.775(6) & 0.71(8) &  28.785(5) & 0.88(15) &  28.790&0.77 &  28.787 &     N VI \\ 
 29.078(9) & 0.24(5) &  29.084(9) & 0.29(7) &  29.082(12) & 0.29(10) &  29.090&0.32 &  29.084 &     N VI \\ 
 29.524(6) & 0.43(6) &  29.520(8) & 0.38(6) &  29.546(11) & 0.43(13) &  29.530&0.41 &  29.534 &     N VI \\ 
 30.445(12) & 0.20(5) &  30.446(13) & 0.22(5) &  30.450(25) & 0.18(11) &  30.448&0.22 &  30.448 &    Ca XI \\ 
 31.027(16) & 0.18(5) &  31.021(12) & 0.15(5) &  31.054(15) & 0.22(10) &  31.015&0.19 &  31.015 &   Si XII \\ 
   ---     & ---     &  33.490(26) & 0.15(7) &  33.510(18) & 0.17(10) &     ---&--- &     --- &  ---       \\ 
 33.724(2) & 3.49(17) &  33.726(2) & 4.15(30) &  33.731(2) & 4.02(32) &  33.736&4.64 &  33.736 &     C VI \\ 
 34.967(12) & 0.19(6) &  34.962(15) & 0.17(7) &  34.959(15) & 0.27(17) &  34.970&0.22 &  34.973 &      C V \\ 
 35.198(30) & 0.13(6) &  35.193(12) & 0.27(8) &  35.188(18) & 0.29(15) &  35.212&0.12 &  35.212 &    Ca XI \\ 
 35.566(16) & 0.15(7) &  35.562(12) & 0.26(7) &  35.566(16) & 0.23(14) &  35.576&0.27 &  35.576 &    Ca XI \\ 
 35.682(8) & 0.37(8) &  35.676(9) & 0.38(8) &  35.672(9) & 0.53(16) &  35.665&0.28 &  35.665 &   S XIII \\ 
 36.374(12) & 0.35(10) &  36.372(15) & 0.28(7) &  36.399(15) & 0.34(14) &  36.398&0.25 &  36.398 &    S XII \\
 36.544(19) & 0.15(6)&  36.561(19)  & 0.29(9)& 36.547(15)   &  0.24(13) & 36.563&0.24& 36.563 & S XII\\

\end{tabular}
\end{center}
\begin{flushleft}
{
\begin{description}
\item $^a$ Identifications for MEKAL (Mewe et al. 1995) and KELLY (Kelly 1987) identical.
\item $^b$ Observed flux in 10$^{-4}$ photons/cm$^2$/s.
\item $^c$ Model flux in 10$^{-4}$ photons/cm$^2$/s, obtained from 3-T fitting of LETGS (see Sect.~3.2).
\item $^d$ Fe XVII lines are strongly mixed with Fe XVI satellite lines.
\item $^e$ Line not split up; mixture of 15.175 \AA\ and 15.265 \AA\ from O VIII and Fe XVII, respectively.
\item $^f$ Small contamination by Fe XVIII possible.\\
Values in parentheses are statistical 1$\sigma$ uncertainties in the last digits.
\end{description}
}
\end{flushleft}
\end{table*}

\section {Observations}
 The spectrum of Procyon was obtained during 70.4 ksec (on November 8, 1999) by 
the High Resolution Camera (HRC-S) and the LETGS on board Chandra. The HRC-S contains three flat detectors, each 10 cm long. 
LETGS consists of 180 grating modules. The LETGS spectrum covers the range from 5 to 175~\AA. 
The LETGS spectra were summed over the +1 and -1 orders and contain also the higher orders. The higher orders are
fitted in the model calculations, but can be neglected for Procyon.
The curve of the effective area as a function of wavelength is complicated
because of the presence of absorption edges (e.g., around 42 \AA) and gaps
between 62--65~\AA\ and 52--56~\AA\ (+1 \& --1 order, respectively) due to
the gaps between the detector plates. 
We use the SRON values (based on calibration by van der Meer et al. 2002), 
which agree within about 5--10\% with the CXC values, as given in
 the Chandra LETGS Calibration Review 
of 31 Oct. 2001.\footnote{\textrm{http://cxc.harvard.edu/cal/Links/Letg/User/Review$_-$311001/\\
eff$_-$area.html}}
  The wavelength resolution is $\Delta\lambda\sim$0.06 \AA\ (FWHM). 
The wavelength uncertainty in the calibration is a few m\AA\ below 30~\AA\ and about 0.020~\AA\ in the region above 30~\AA.
The spectra are background subtracted. 
The statistical errors in the line fluxes include errors from the background.
For further instrumental details see also Brinkman et al. (2000).

Later (on October 22, 2000), the spectrum of Procyon was observed by XMM-Newton using 
the RGS and EPIC-MOS. The total observing time was $\approx$~107~ksec; however, due to
large solar flare activity at the end of the observation, we removed 16.7~ksec of data, leaving a total of 90.5~ksec of
``good'' data. In XMM-Newton three telescopes focus X-rays onto three EPIC cameras (two MOS and one pn). About half 
of the photons in the beams of two telescopes (Turner et al. 2001) are diffracted by 
sets of reflecting gratings and are then focussed onto the RGS detectors. The 
RGS spectral resolution is $\Delta\lambda\sim$0.07 \AA, with a maximum
effective area of about 140 cm$^2$ around 15 \AA. The wavelength uncertainty is 7-8 m\AA.
The first spectral order has been selected by means of the energy resolution of the individual CCDs.
For further details see den Herder et al. (2001).

The data were processed by the XMM-Newton SAS using the calibration of February 2001.
 The RGS cover the range from 5 to 37 \AA. 
The EPIC spectra, which have a lower resolution but higher sensitivity, are used to constrain the high-temperature part 
of Procyon's EM distribution.
 Because of the high resolution of the grating spectrometers we will focus on the spectra from these instruments.
 In Fig.~1, we show the RGS spectra together with an extract of the LETGS spectrum covering the 
 wavelength range from 10 to 37~\AA. No notable features are observed below 10 \AA\ in the LETGS and RGS spectra. However, the EPIC-MOS detects
 the H- and He-like lines of Mg. The remaining part of the LETGS spectrum is shown in Fig.~2. 
 From Fig.~1 the gaps in the two RGS spectra due to CCD failure of CCD 7 (RGS1) and 4 (RGS2) are obvious.

\section{Analysis}
\subsection{Detailed analysis of the spectra}
The spectral lines from all three instruments have been measured individually. We folded monochromatic delta functions through the instrumental 
response matrices in order to derive the integrated line fluxes. No additional width was needed to fit the shape of the lines. A constant "background" level was 
adjusted in order to account for the real continuum and for the pseudo-continuum created by the overlap of several 
weak, neglected lines.
In Table~1, we have collected the measured wavelengths and fluxes of the emission lines in the 
RGS instruments together with those in the LETGS in the similar wavelength range. 
The fluxes among the three data sets, as collected in Table~1, are in good agreement in view of the systematic uncertainties
in the calibration. However for some lines deviations appear, which are 
caused by gaps between the individual CCDs.
In this wavelength range (below 40~\AA) the identification is in general straightforward. The dominating lines are strong and  
belong to H- and He-like ions for which atomic parameters are well known. Although XMM-Newton and Chandra observed Procyon at different dates no strong differences in flux (Table~1) are noticed, resulting in the
conclusion that the coronal emission of Procyon did not vary strongly from one observation to the other. 
 
Table~2 contains the same information as Table~1 for LETGS lines which occur above 37~\AA. 
The extracted fluxes are as measured at Earth. Therefore they are not corrected for interstellar absorption which is of the order
of 4\% at 100~\AA, 6\% at 125~\AA, 10\% at 150~\AA, and 15\% at 175~\AA.
We added one Fe line (Fe~X at 174.69~\AA) that was observed in an offset observation of Procyon (obsID = 1224; 14.8~ksec). For that line 
the effective area was obtained by extrapolation. The line flux ratio of that line compared to the line at 171.075~\AA\ in 
the offset observation was used to establish the flux value.

In both tables, we have compared the measured wavelengths 
with the wavelengths in various atomic databases: 
the MEKAL (Mewe et al. 1985, 1995) code\footnote{\textrm{http://saturn.sron.nl/$\sim$kaastra/spex/line.ps.gz}}, KELLY (Kelly 1987) and the
database of the National Institute of Standards and Technology (NIST), which is also available 
on the web.\footnote{\textrm{http://physics.nist.gov/cgi-bin/AtData/main$_-$asd}}
We have also compared with a list of lines observed in the solar corona (Doschek \& Cowan 1984, hereafter D\&C). 
Further we compare our measured iron lines with the results
from laboratory experiments such as the Lawrence Livermore National Laboratory's
Electron Beam Ion Trap (EBIT) (see Beiersdorfer et al. 1999 and Lepson et al. 2002 for Fe VIII--X and Lepson et al. 2000 for 
Fe XII--XIII).
A number of lines in Table~2 (see note "d") are in close wavelength agreement to lines identified in EBIT.
Finally in Table~1 the fluxes, from the multi-temperature global fitting of Sect.~3.2, have been added.

Some possible line identifications have been omitted from Table~2, due to the absence of comparable lines belonging to the same multiplet or ion 
(Table~3) or due to ambiguity of the identification of lines in atomic databases (Kelly 1987). The latter concerns lines at 60.989~\AA\ 
(Si VII, VIII, \& IX) and 61.852 \AA\ (Si VIII \& IX).

\begin{table}
\caption{Wavelengths and fluxes for LETGS above 36.5 \AA, together with
the line identifications from MEKAL, KELLY, and D\&C}         
 \begin{tabular}{|l@{\ }l@{}|l@{\ }r@{\ }|l@{\ }r@{\ }|l@{\ }r@{ }|}
\hline 
\multicolumn{2}{|c|}{LETGS}&\multicolumn{6}{c|}{Line identifications$^a$}\\
\hline
\multicolumn{2}{|c|}{}&\multicolumn{2}{c|}{MEKAL}&\multicolumn{2}{c|}{KELLY}&\multicolumn{2}{c|}{D\&C}\\
\hline
$\lambda$(\AA) &  flux$^b$&$\lambda$(\AA) & Ion&$\lambda$(\AA) &Ion&$\lambda$(\AA) &  Ion   \\
\hline
39.276  &0.63(14) & 39.300 &  S XI & 39.300 &  S XI & 39.30 &  S XI \\ 
   &    &        &      &39.264&Si X&  &\\
   &    &        &      &39.305&Si X&  &\\
40.263 & 2.29(36) & 40.270 &   C V & 40.268 &   C V & 40.27 &   C V \\ 
40.718 & 1.88(42) & 40.730 &   C V & 40.731 &   C V & 40.73 &   C V \\ 
41.475 & 1.07(29) & 41.470 &   C V & 41.472 &   C V & 41.47 &   C V \\ 
       &      & 41.480 & Ar IX & 41.480 & Ar IX &   &       \\
42.543 & 1.29(28) & 42.530 &   S X & 42.543 &   S X & 42.53 &   S X \\ 
42.810 & 0.33(17) & --&    & 42.826&Si XI& -- &      \\
43.743 & 0.54(8) & 43.740 & Si XI & 43.763 & Si XI & 43.74 & Si XI \\ 
44.014 & 0.43(8) & 44.020 & Si XII & 44.021 & Si XII & 44.02 & Si XII \\ 
44.150 & 0.67(10) & 44.165 & Si XII & 44.165 & Si XII & 44.17 & Si XII \\ 
44.218 & 0.52(10) & 44.249 & Si IX & 44.215 & Si IX & 44.22 & Si IX \\ 
45.677 & 0.20(4) & 45.680 & Si XII & 45.692 &  Si XII & 45.68 & Si XII \\ 
   &    &    &      & 45.684 & Si X   &    &      \\ 
46.283 & 0.25(7) & 46.300 & Si XI & 46.300 & Si XI & 46.30 & Si XI \\ 
46.391 & 0.40(8) & 46.410 & Si XI & 46.401 & Si XI & 46.41 & Si XI \\ 
47.242 & 0.46(8) & 47.280 &  Mg X & 47.310 &  Mg X & 47.31 &  Mg X \\ 
   &    &    &      & 47.231 &  Mg X &    &    \\ 
   &    &    &      & 47.249 &  S IX & 47.25 &  S IX \\ 
47.452 & 0.48(8) & 47.500 &  S IX & 47.433 &  S IX & 47.43 &  S IX \\ 
   &    &    &      & 47.518 &  S IX &    &      \\ 
   &    &    &      & 47.453 &  Si XI &    &      \\   
47.642 & 0.49(8) & 47.654 &   S X & 47.655 &  S X & 47.65 &   S X \\ 
       &      &        &      &47.653& Si XI &  &  \\
47.774 & 0.34(7) & 47.793 & S X    & 47.791 & S X    & 47.79 &  S X  \\ 
47.883 & 0.24(8) & 47.896 &  Mg X & 47.905 &   S X & 47.90 &  Mg X \\ 
       &      &        &      & 47.899 & Si XI & & \\
48.720 & 0.23(6) & 48.730 & Ar IX & 48.73  & Ar IX & 48.73 & Ar IX \\ 
49.109 & 0.33(8) & -- &      & 49.119 &  S IX & 49.12 &  S IX \\ 
49.207 & 1.44(14)   & 49.220 & Si XI & 49.222 & Si XI & 49.22 & Si XI \\ 
       &      & 49.180 & Ar IX & 49.18  & Ar IX & 49.18 & Ar IX \\ 
49.324 & 0.32(7) & -- &      & 49.328 &  S IX &--    &      \\ 
49.696 & 0.29(7) & -- &      & 49.701 &  Si X &--    &      \\ 
49.975 & 0.28(6)   & 50.019 & Si VIII$^c$ &50.019    &Si VIII      &--    &      \\ 
50.327 & 0.51(8) & -- &      & 50.333 & Si X & -- &       \\ 
50.520 & 1.68(15) & 50.530 &  Si X & 50.524 &  Si X & 50.53 &  Si X \\ 
50.686 & 1.30(14) & 50.690 &  Si X & 50.691 &  Si X & 50.69 &  Si X \\ 
52.306 & 0.75(11) & 52.300 & Si XI & 52.296 & Si XI & 52.30 & Si XI \\ 
52.594 & 0.35(8) & 52.615 & Ni XVIII & 52.615 & Ni XVIII & -- &      \\ 
   &    &    &      & 52.611 & Si IX    &    &      \\ 
52.715 & 0.30(7) & 52.720 & Ni XVIII & 52.720 & Ni XVIII & 52.70 & S VIII \\ 
52.772& 0.30(7)&--  &      & 52.756 & S VIII &--  &      \\ 
      &     &    &      &52.789  & S VIII &    &    \\
52.898 & 0.30(7) &52.911 & Fe XV  & 52.911 & Fe XV  & 52.87 & Fe XV  \\
54.133 & 0.56(13) & 54.118 & S VIII & 54.118 & S VIII & 54.12 & S VIII \\ 
   &    & 54.142 & Fe XVI & 54.142 & Fe XVI & 54.15 & Fe XVI \\ 
54.180 & 0.31(10) & 54.180 &  S IX & 54.175 &  S IX & 54.18 &  S IX \\ 
54.546 & 0.54(13) & -- &      & 54.571 &  Si X & -- &      \\ 
   &    &    &      & 54.566 &S VIII &    &      \\ 
54.700 & 0.45(11)& 54.728 & Fe XVI & 54.728 & Fe XVI & 54.70 & Fe XVI \\ 
55.094 & 0.68(15) & 55.060 & Mg IX & 55.060 & Mg IX & 55.06 & Mg IX \\ 
   &    &    &      & 55.094 & Si IX & 55.12 & Si IX \\ 
   &    &    &      & 55.116 & Si IX &    &      \\ 
   &    &    &      & 55.096 & Si X &    &      \\ 
55.270&0.88(25)&55.272 &Si IX    & 55.272 & Si IX & 55.27 & Si IX \\ 
   &    &    &      & 55.305 & Si IX & 55.31 & Si IX \\  
\hline
\end{tabular}
\end{table}

\begin{table} 
 \begin{tabular}{|l@{\ }l@{}|l@{\ }r@{\ }|l@{\ }r@{\ }|l@{\ }r@{ }|}
\hline 
\multicolumn{2}{|c|}{LETGS }&\multicolumn{6}{c|}{Line identifications$^a$}\\
\hline
\multicolumn{2}{|c|}{}&\multicolumn{2}{c|}{MEKAL}&\multicolumn{2}{c|}{KELLY}&\multicolumn{2}{c|}{D\&C}\\
\hline
$\lambda$(\AA) &  flux$^b$&$\lambda$(\AA) & Ion&$\lambda$(\AA) &Ion&$\lambda$(\AA) &  Ion   \\
\hline
55.359 & 2.14(27) & 55.356 & Si IX & 55.356 & Si IX & 55.36 & Si IX \\ 
   &    &    &      & 55.401 & Si IX & 55.40 & Si IX \\ 
56.037 &0.19(11)& 56.000 & Ni XIII & 56.027 & Si IX & 56.03 & Si IX \\ 
   &    &    &      & 56.081 &  S IX & 56.08 &  S IX \\ 
56.836   &0.20(8)    & -- &      & 56.804 & Si X & -- &      \\
57.741 & 0.80(35)  & --    &      & 57.736&Mg VIII&-- &  \\
       &       &       &      & 57.778& Si IX &   &  \\
57.856 &0.78(35)    & 57.880 &  Mg X & 57.876 &  Mg X & 57.88 &  Mg X \\ 
   &    & 57.920 &  Mg X & 57.920 &  Mg X & 57.92 &  Mg X \\ 
61.020 & 1.41(25) & 61.050 & Si VIII & 61.019 & Si VIII & 61.01 & Si VIII \\ 
   &    &    &      & 61.038 & Mg IX &    &      \\ 
61.087 & 1.38(24) & -- &      & 61.070 & Si VIII & 61.08 & Si VIII \\ 
       &      &    &      & 61.088 & Mg IX   &        &        \\ 
61.578 & 0.52(17) & 61.600 & S VIII& 61.600 & S VIII& 61.60 & S VIII \\ 
61.668 & 0.49(15) &--      &       & 61.649 & Si IX & 61.66 & Si IX \\ 
61.843 & 0.67(11) & 61.841 & Si IX & 61.852 & Si IX & 61.84 & Si IX \\ 
61.916 & 0.55(17) & 61.912 & Si VIII & 61.914 & Si VIII & 61.91 & Si VIII \\ 
   &    &    &      & 61.895 & Si VIII & 61.90 & Si VIII \\ 
62.748 & 0.53(17) & 62.755 &Mg IX  & 62.751 &Mg IX  & 62.76 & Mg IX \\ 
   &    & 62.699 &Fe XIII$^d$& 62.694 & Fe XIII &    &      \\ 
62.849 & 0.38(11) & 62.879 &Fe XVI & 62.879 &Fe XVI & 62.88 & Fe XVI \\ 
       &      & 62.800 & Fe X$^d$ &62.8   &Fe X &       &      \\
63.161 & 0.64(13) & 63.153 &  Mg X & 63.152 &  Mg X & 63.15 &  Mg X \\ 
63.283 & 0.94(15) & 63.294 &  Mg X & 63.295 &  Mg X & 63.29 &  Mg X \\ 
63.390 & 0.38(8) & 63.314 &  Mg X & 63.304 & S VIII & 63.40 & Mg VII \\ 
       &      &        &      & 63.396 & Mg VII&  &  \\
63.720 & 0.58(11) & 63.719 & Fe XVI & 63.719 & Fe XVI & 63.71 & Fe XVI \\ 
   &    &        &       & 63.732 & Si VIII &63.73 &Si VIII      \\ 
63.921 & 0.39(10) & --   &    &--    &     & -- &      \\
64.135  & 0.44(11)   & --   &    &--    &     & -- &      \\
65.677 & 0.38(11) & 65.672 &  Mg X & 65.672 &  Mg X & 65.67 &  Mg X \\ 
65.826 &0.49(13)& 65.840 &  Mg X &65.847&Mg X& 65.84 &  Mg X \\
       &      &         &     &65.822&Ne VIII   &  &  \\
65.884   &0.41(10)    & 65.905 & Fe XII$^d$ &   65.905 & Fe XII & --    &      \\ 
         &        &        &       & 65.892 & Ne VIII&   --&        \\ 
66.057 & 0.28(10) & 66.047 & Fe XII$^d$ & 66.047 & Fe XII & 66.04 & Fe XII \\ 
66.255 & 0.64(14) & --  &    & 66.259 & Ne VIII & --  &    \\ 
66.352  &  0.63(13)   &  --        &      & 66.330 & Ne VIII &  --   &    \\
67.161 & 0.48(11) & 67.132 &Mg IX&67.135   &Mg IX & 67.13&Mg IX\\
67.255 & 0.87(18) & 67.233 & Mg IX & 67.239 & Mg IX & 67.22 & Mg IX \\ 
       &      &        &      &67.291&Fe XII$^d$   &    &  \\
67.375 & 0.68(14) & 67.350 & Ne VIII & 67.382 & Ne VIII & 67.35 & Ne VIII \\ 
69.646 & 2.03(21) & 69.658 & Si VIII & 69.632& Si VIII     & 69.66 & Si VIII \\ 
       &      & 69.660 & Fe XV   & 69.66 & Fe XV       &        &         \\ 
69.827 & 1.05(14) & 69.825 & Si VIII & 69.790& Si VIII & 69.83 & Si VIII \\ 
70.046 & 0.70(11) & 70.020 &Si VII & 70.027&Si VII & 70.03 & Si VII \\ 
   &    & 70.010 & Fe XII& 70.01  & Fe XII&  & \\ 
   &    & 70.054 &Fe XV    & 70.054 &Fe XV   &70.05  & Fe XV      \\ 
71.929  &0.69(13 & --   &      & 71.901 & Mg IX & 71.92   & Mg IX     \\ 
     &      &    &      & 71.955 & Si VII &  &  \\ 
72.034   &0.43(13)    & 72.030 & Mg IX & 72.027 & Mg IX & 72.03 & Mg IX \\ 
72.302 & 1.44(18) & 72.311 & Mg IX & 72.312 & Mg IX & 72.31 & Mg IX \\ 
       &      & 72.310 & Fe XI & 72.310 & Fe XI &        &       \\ 
72.668&0.73(14) & 72.663 & S VII & 72.663 & S VII & 72.66 & S VII \\ 
72.871 & 0.58(15) & 72.850 & Fe IX$^d$ & 72.850 & Fe IX & -- &      \\ 
73.478 & 0.47(11) & -- &  & 73.470 & Ne VIII & -- &  \\ 
   &    & 73.471  & Fe XV & 73.471 & Fe XV & 73.47 & Fe XV  \\
73.555   &0.43(11)    & 73.560 & Ne VIII & 73.563 & Ne VIII & 73.56 & Ne VIII \\ 
74.860 & 1.10(18) & 74.854 & Mg VIII & 74.858 & Mg VIII & 74.85 & Mg VIII \\ 
       &      & 74.845 & Fe XIII$^d$&74.845&Fe XIII&  &  \\ 
    
\hline
\end{tabular}
\end{table} 

\begin{table} 
 \begin{tabular}{|l@{\ }l@{}|l@{\ }r@{\ }|l@{\ }r@{\ }|l@{\ }r@{ }|}
\hline 
\multicolumn{2}{|c|}{LETGS }&\multicolumn{6}{c|}{Line identifications$^a$}\\
\hline
\multicolumn{2}{|c|}{}&\multicolumn{2}{c|}{MEKAL}&\multicolumn{2}{c|}{KELLY}&\multicolumn{2}{c|}{D\&C}\\
\hline
$\lambda$(\AA) &  flux$^b$&$\lambda$(\AA) & Ion&$\lambda$(\AA) &Ion&$\lambda$(\AA) &  Ion   \\
\hline
75.035 & 1.05(18) & 75.034 & Mg VIII & 75.034 & Mg VIII & 75.03 & Mg VIII \\ 
75.978 & 0.47(11) & 76.006 &  Fe X$^d$ & 76.006 &  Fe X & 76.02 & Fe X \\ 
76.038   & 0.77(13)   &  -- &      & -- &      & -- &      \\
76.507 & 0.32(11) & 76.502 & Fe XVI & 76.502 & Fe XVI & 76.51 & Fe XVI \\ 
76.862 & 0.55(13) & -- &      & -- &      & 76.87 & Fe XVI \\
77.740 & 1.11(18) & 77.741 & Mg IX & 77.737 & Mg IX & 77.74 & Mg IX \\ 
78.733 & 0.71(14) & 78.717 & Ni XI & 78.744 & Ni XI & 78.72 & Ni XI \\ 
        &      & 78.769 &  Fe X$^d$ & 78.769 &  Fe X &    &      \\ 
79.483 & 0.58(13) & 79.488 & Fe XII & 79.488 & Fe XII & 79.49 & Fe XII \\ 
80.017 & 0.38(11) & 80.022 & Fe XII & 80.022 & Fe XII & 80.02 & Fe XII \\ 
80.236   & 0.54(14)   & --   &      & 80.255 & Mg VIII & --   &      \\
80.507 & 0.74(14) & 80.501 & Si VI & 80.501 & Si VI & 80.50 & Si VI \\ 
   &    & 80.510 & Fe XII & 80.510 & Fe XII & 80.51 & Fe XII \\ 
80.751 & 0.54(14) & -- &      & 80.725 & Si VI & -- &      \\ 
81.865 &0.42(11) & --&  & 81.895 & Si VII & -- &      \\ 
82.420 & 0.48(11) & 82.430 & Fe IX$^d$ & 82.430 & Fe IX & 82.43 & Fe IX \\ 
82.667 & 0.97(17) & 82.744 & Fe XII & 82.598 & Mg VIII & -- &      \\
82.808  & 0.35(10)   & 82.837 & Fe XII & 82.837 & Fe XII  & -- &      \\ 
   &    &    &      & 82.822 & Mg VIII&    &      \\ 
83.337 & 0.38(11) & --      &      & 83.358 & Si VI & --   &      \\ 
83.600 & 0.55(15) & -- &      & 83.587 & Mg VII & 83.59 & Mg VII \\ 
   &    &    &      & 83.611 & Si VI &    &      \\ 
83.764 & 0.47(17) & 83.766 & Mg VII & 83.766 & Mg VII & 83.77 & Mg VII \\ 
83.935 & 0.46(11) & 83.959 & Mg VII & 83.959 & Mg VII & 83.96 & Mg VII \\ 
   &    &    &      & 83.910 & Mg VII & 83.91 & Mg VII \\ 
84.032 & 0.40(11) & -- &      & 84.025 & Mg VII & 84.02 & Mg VII \\ 
84.292   & 0.40(11)   & 84.292 & Ne VII & 84.292 & Ne VII & 84.29 & Ne VII \\ 
   &    &    &      & 84.212 & Ne VII &    &      \\ 
84.433 & 0.39(11) & --   &     & --   &    & -- &      \\  
85.448 & 0.38(11) & -- &     & 85.477 &Fe XII  & 85.47 &Fe XII \\ 
   &    &    &      & 85.407 & Mg VII & 85.41 & Mg VII \\ 
86.765 & 1.13(17) & 86.772 & Fe XI & 86.772 & Fe XI & 86.77 & Fe XI \\ 
86.876 & 0.55(17) & -- &      & 86.847 & Mg VIII & -- &      \\
87.021   &0.46(14)    & 87.025 & Fe XI & 87.025 & Fe XI   & 87.02 & Fe XI \\ 
   &    &    &      & 87.017 &Mg VIII&    &      \\ 
88.087  & 1.68(20)   &88.092 & Ne VIII & 88.092 & Ne VIII  & 88.08 &Ne VIII      \\
88.893&0.68(14)&--&  &-- &  &-- &  \\ 
88.955 & 0.75(15) & --     &       & 88.952 & Mg VI & -- &      \\ 
89.156 & 0.43(13) &  89.185 & Fe XI & 89.185 & Fe XI & 89.18 & Fe XI \\ 
90.719 & 0.59(13) & -- &      & 90.708 & Mg VII & -- &      \\ 
90.989 & 0.43(10) & 91.009 & Fe XIV &91.009 & Fe XIV & -- &      \\ 
   &    &    &      & 90.955 & Fe XVII$^e$&    &      \\ 
91.529 & 0.52(13) & 91.564 & Ne VII & 91.564 & Ne VII & 91.56 & Ne VII \\ 
91.627 & 0.38(8) & -- &      &--    &    &--    &      \\ 
91.777 & 0.58(13) & 91.808 &  Ni X & 91.790 &  Ni X & 91.81 &  Ni X \\ 
92.155 & 0.51(13) & -- &      & 92.123 & Mg VIII & -- &      \\ 
92.858 & 0.55(14) & -- &      & 92.850   & Ne VII & -- &      \\ 
93.587   &0.58(15)    & --&  &-- &  &-- &  \\ 
94.001 & 1.70(24) & 94.012 &  Fe X$^d$ & 94.012 &  Fe X & 94.02 &  Fe X \\ 
95.339 & 0.90(18) & 95.374 &  Fe X & 95.374 & Fe X & 95.37 &  Fe X \\ 
   &    & 95.338 &  Fe X$^d$ & 95.338 &  Fe X & -- &    \\ 
95.412 & 1.04(20) & 95.483 & Mg VI & 95.483 & Mg VI & 95.48 & Mg VI \\ 
   &    &    &      & 95.421 & Mg VI & 95.42 & Mg VI \\ 
95.997 & 1.46(20) & -- &      & 96.022 & Si VI & 96.02 & Si VI \\ 
96.124 & 0.79(17) & 96.122 &  Fe X$^d$ & 96.122 &  Fe X & 96.12 &  Fe X \\ 
96.804 & 0.71(18) & 96.788 &  Fe X$^d$ & 96.788 &  Fe X & -- &      \\
97.104 &0.34(15)& 97.122 &  Fe X$^d$ & 97.122 &  Fe X & 97.12 &  Fe X \\ 
97.486 & 0.78(17) & 97.502 & Ne VII & 97.502 & Ne VII & 97.50 & Ne VII \\ 
98.091 & 1.58(25) & 98.115 & Ne VIII & 98.115 & Ne VIII & 98.13 & Ne VIII \\ 
98.251 & 2.89(34) & 98.260 & Ne VIII & 98.260 & Ne VIII & 98.26 & Ne VIII \\ 
\hline
\end{tabular}
\end{table} 

\begin{table} 
 \begin{tabular}{|l@{\ }l@{}|l@{\ }r@{\ }|l@{\ }r@{\ }|l@{\ }r@{ }|}
\hline 
\multicolumn{2}{|c|}{LETGS }&\multicolumn{6}{c|}{Line identifications$^a$}\\
\hline
\multicolumn{2}{|c|}{}&\multicolumn{2}{c|}{MEKAL}&\multicolumn{2}{c|}{KELLY}&\multicolumn{2}{c|}{D\&C}\\
\hline
$\lambda$(\AA) &  flux$^b$&$\lambda$(\AA) & Ion&$\lambda$(\AA) &Ion&$\lambda$(\AA) &  Ion   \\
\hline
100.57 & 1.05(24) & -- &      & 100.597 & Mg VIII &  -- &  \\ 
102.85 & 0.90(22) & 102.91 & Ne VIII & 102.911 & Ne VIII & 102.9 & Ne VIII \\ 
103.07 & 1.68(27) & 103.08 & Ne VIII & 103.085 & Ne VIII & 103.1 & Ne VIII \\ 
103.54 & 2.08(32) & 103.57 & Fe IX$^d$ & 103.566 & Fe IX & 103.6 & Fe IX \\ 
103.88 & 0.72(22) & -- &      & --  &    & --   &      \\
104.67 & 0.68(21) & -- &      &  -- &     & -- &      \\ 
104.78 & 0.86(21) & 104.81 &  O VI & 104.813 &  O VI & -- &      \\ 
105.20 & 1.22(21) & 105.21 & Fe IX$^d$ & 105.208 & Fe IX & 105.2 & Fe IX \\ 
106.18 & 1.03(20) & 106.19 & Ne VII & 106.192 & Ne VII & 106.2 & Ne VII \\
111.23 & 1.49(28) & -- &      & 111.198 &  Ca X & --   &      \\
111.71   &0.53(13)    & 111.57 & Mg VI & 111.552 & Mg VI & 111.6 & Mg VI \\ 
   &    & 111.72 & Mg VI & 111.746 & Mg VI & 111.7 & Mg VI \\ 
113.33 & 0.46(11) & -- &      & 113.315 & Fe VIII & --   &      \\ 
113.77&0.60(20)& -- &      &113.763    &Fe VIII    &--    &   \\
      &    &    &      &113.793   &Fe IX&   & \\
113.99 & 0.71(20) & -- &      & 113.990 &  Mg V & 114.0 &  Mg V \\ 
   &    &    &      & 114.029 &  Mg V &    &      \\ 
114.54 & 0.48(14) & -- &      & 114.564 & Fe VIII & -- &      \\ 
114.88 & 0.66(17) & -- &      & 114.785 &  Mg V & 114.8 &  Mg V \\ 
115.37   &0.51(18)    & 115.33 & Ne VII & 115.33 & Ne VII & -- &    \\ 
         &        & -      &       & 115.39 & Ne VII & -- &    \\ 
115.80 & 0.77(20) & 115.83 &  O VI & 115.826  &  O VI & 115.8 &  O VI \\ 
115.89&0.46(17)& -- &      & --    &      &--    &   \\
116.70 & 1.54(25) & 116.69 & Ne VII & 116.693 & Ne VII & 116.7 & Ne VII \\ 
116.87&0.54(14)& -- &      & -- &      & -- &      \\
117.20   & 0.54(20)   & 117.20 & Fe VIII$^d$ & 117.197 & Fe VIII & -- &      \\ 
117.66 & 0.80(20) & -- &      & -- &      & -- &      \\  
119.31 &0.46(13)& -- &      & -- &      & -- &      \\  
120.31  &0.60(15)& 120.33 & O VII & 120.331 & O VII & -- &      \\ 
122.49 & 0.73(18) & 122.49 & Ne VI & 122.49  & Ne VI & 122.5 & Ne VI \\ 
123.54 & 1.31(31) & -- &      & --      &      & -- &      \\ 
124.51 & 0.98(25) & -- &      & -- &      & -- &      \\ 
126.25 & 0.98(32) & -- &      & 126.280 &  Mg V & -- &      \\ 
127.53 & 0.98(25) & -- &      & --      &      & -- &      \\ 
127.69 & 1.44(29) & 127.66 & Ne VII & 127.663 & Ne VII & 127.7 & Ne VII \\ 
129.86 & 1.07(34) & 129.83 &  O VI & 129.871 &  O VI & 129.9 &  O VI \\ 
130.92 & 1.78(39) & 130.94 & Fe VIII$^d$ & 130.941 & Fe VIII & 130.9 & Fe VIII \\ 
131.21 & 1.78(35) & 131.24 & Fe VIII$^d$ & 131.240 & Fe VIII & 131.2 & Fe VIII \\ 
134.21 & 1.43(41) & -- &      &  --     &       & -- &      \\ 
135.48 & 0.97(35) & 135.52 &   O V & 135.523 &   O V & 135.5 &   O V \\ 
136.78 & 1.91(53) & -- &      & -- &      & -- &      \\ 
140.27 & 1.36(49) & -- &      & --      &      & -- &      \\ 
141.04 & 0.79(22) & 141.04 & Ca XII & 141.038 & Ca XII & 141.0 & Ca XII \\ 
144.97 & 1.73(49) & 144.99 &  Ni X & 144.988 &  Ni X & 145.0 &  Ni X \\ 
147.27&1.12(35)& 147.27 & Ca XII     & 147.278 &Ca XII      & 147.3 &Ca XII    \\
148.36 & 11.3(10)  & 148.40 & Ni XI & 148.402 & Ni XI & 148.4 & Ni XI \\ 
150.08 & 5.08(60) & 150.10 &  O VI & 150.1   &  O VI & 150.1 &  O VI \\ 
151.52 & 2.36(39) & -- &      & 151.548 &   O V & -- &      \\ 
152.11 & 5.60(66) & 152.15 & Ni XII & 152.153 & Ni XII & 152.2 & Ni XII \\ 
154.14 & 2.73(42) & 154.18 & Ni XII & 154.175 & Ni XII & 154.2 & Ni XII \\ 
155.56 & 1.36(28) & -- &      & -- &      & -- &      \\ 
156.14 & 1.66(32) & -- &      & 156.140 &  Ne V & -- &      \\ 
156.38   &1.29(31)    & -- &      & -- &      & -- &      \\ 
157.68 & 2.76(42) & 157.73 & Ni XIII & 157.730 & Ni XIII & 157.7 & Ni XIII \\
158.33 & 2.08(35) & 158.38 &  Ni X & 158.377 &  Ni X & 158.4 &  Ni X \\
158.78 & 1.47(34) & -- &      & 158.770 & Ni XIII & -- &      \\ 
159.24 & 1.03(27) & -- &      & 159.300 &  Si X & 159.1 & Ar XIII \\ 
159.58 & 1.83(43) & -- &      &    --   &      & -- &      \\ 
159.93 & 3.09(43) & 159.94 &  Ni X & 159.977 &  Ni X & 159.9 &  Ni X \\ 
   &    & 159.97 & Ni XIII & 159.97  & Ni XIII &    &      \\ 
 
\hline
\end{tabular}
\end{table} 

\begin{table} 
 \begin{tabular}{|l@{\ }l@{}|l@{\ }r@{\ }|l@{\ }r@{\ }|l@{\ }r@{ }|}
\hline 
\multicolumn{2}{|c|}{LETGS }&\multicolumn{6}{c|}{Line identifications$^a$}\\
\hline
\multicolumn{2}{|c|}{}&\multicolumn{2}{c|}{MEKAL}&\multicolumn{2}{c|}{KELLY}&\multicolumn{2}{c|}{D\&C}\\
\hline
$\lambda$(\AA) &  flux$^b$&$\lambda$(\AA) & Ion&$\lambda$(\AA) &Ion&$\lambda$(\AA) &  Ion   \\
\hline 
162.56 & 2.94(83) & 162.56 &   N V & 162.556 &   N V & -- &      \\ 
164.11 & 3.53(74) & 164.15 & Ni XIII & 164.146 & Ni XIII & 164.1& Ni XIII \\ 
167.43 & 3.9(10) & 167.49 & Fe VIII & 167.486 & Fe VIII & 167.5 & Fe VIII \\ 
167.59 & 3.9(12) & 167.66 & Fe VIII & 167.656 & Fe VIII & -- &      \\ 
168.13 & 7.4(13) & 168.17 & Fe VIII & 168.172 & Fe VIII & 168.2 & Fe VIII \\ 
168.51 & 5.4(13) & 168.54 & Fe VIII & 168.545 & Fe VIII & 168.5 & Fe VIII \\ 
168.90 & 5.0(18) & 168.93 & Fe VIII & 168.929 & Fe VIII & 168.9 & Fe VIII \\ 
171.04 & 114(8) & 171.08 & Fe IX & 171.075 & Fe IX & 171.1 & Fe IX \\ 
174.49 & 118(24) & 174.53 & Fe X & 174.53 & Fe X & 174.5 & Fe X \\ 
\hline
\end{tabular}
\begin{flushleft}
{
\begin{description}
\item $^a$ from MEKAL (Mewe et al. 1995), KELLY (1987), and D\&C (solar
line list of Doschek and Cowan 1984). 
\item $^b$ Observed flux in 10$^{-4}$ photons/cm$^2$/s with in parentheses
1$\sigma$ uncertainty in the last digits.
\item $^c$ MEKAL placed it at 52.0 \AA.          
\item $^d$ line identified in EBIT spectrum.
\item $^e$ from CHIANTI (Dere et al. 1997).
\end{description}
}
\end{flushleft}
\end{table}

\begin{table}
\caption{Possible line identifications left out of Table~2. Col.~$\lambda$: observed
wavelengths from Table 2. Cols.~2 and 3 give a possible identification which 
has not been given in Table 2, due to the absence of the lines, given in Cols.~4 and 5}
\begin {center}
\label{upper1}
\begin{tabular}{l|rlrl}
$\lambda$(\AA)  &  present  &  ion  &  missing & ion \\
\hline
93.587        &  93.616   & FeVIII& 93.469   & FeVIII\\
              &           &       & 108.077  & FeVIII\\
98.583        &  98.548   & FeVIII&  98.371  & FeVIII\\       
103.88        &  103.937  & FeXVIII& 93.923  & FeXVIII\\
103.88        &  103.904  & MgV    &110.859  & MgV \\

\end{tabular}
\end{center}
\end{table}

Earlier benchmarks with a solar flare spectrum (Phillips et al. 1999) and with RGS and LETGS spectra of Capella
(Audard et al. 2001a; Mewe et al. 2001) have already shown that the 
current atomic databases are lacking quite a number of spectral lines 
 for L-shell transitions of Ne, Mg, Si, and S, that appear
in the long-wavelength region above about 40 \AA. This is illustrated by the many identifications present in the third 
Col.~(KELLY), which are absent in MEKAL. For the Fe L-shell Behar et al. (2001) have shown that the HULLAC atomic data 
 are fairly accurate.

\subsection{Global Fitting and emission measure modeling}
\subsubsection{Multi-temperature fitting}
We first characterize the thermal structure and the elemental composition of Procyon's corona. To this end, we fitted
multi-T models using SPEX (Kaastra et al. 1996a) of the spectra (RGS+MOS and LETGS). For both the observations the 
calculations require two dominant temperature components. A third (small and not very significant) temperature 
component is needed to account for the lines of low stages of ionization, present in the LETGS spectrum.
The reduced $\chi^2$ is relatively high (1.3--2) for the fits. This is 
due to a lack of lines in the MEKAL code and to the high resolution of the instrument. Small wavelength deviations 
(about 1-2 bins i.e. 0.02-0.04 \AA) between 
lines in the spectrum and in the model are often present (see Table~2). This effect results in a sharp maximum and minimum 
in the value of the normalized difference between model and observation around the peak of the line (see also Fig.~4).
The results of RGS and LETGS are very similar.\\ 
In Table 4 results for temperatures $T$ (in MK), emission measures $EM$, and abundances are given.
Statistic 1$\sigma$ uncertainties are given in parentheses.
The emission measure is defined as $EM = n_\mathrm{e} n_\mathrm{H} V$, where $V$ is the volume contributing 
to the emission and for solar abundances the hydrogen density $n_\mathrm{H} \simeq 0.85 n_\mathrm{e}$.
The temperatures and emission measures of all spectra show a dominant region between 1 and 2.5 MK. The two dominant temperature 
components are about 1.2 and 2.3~MK. Using EUVE, Schmitt et al. (1996) derived a DEM with a peak temperature of 1.6 MK based on 
Fe-lines only. This is in satisfactory agreement with our results.

The total emission measures summed over all temperature components 
are about $4.6(.4) \times 10^{50}$~cm$^{-3}$ for LETGS and $3.9(.3) \times 10^{50}$~cm$^{-3}$ for RGS+MOS. 
These are similar to the
 total $EM$ of $4.5 \times 10^{50}$~cm$^{-3}$ found by Schmitt et al. (1996). 

The determination of abundances is complicated by several factors. The many weak L-shell lines, which are absent 
in the atomic code (see difference between Col.~"MEKAL" and "KELLY" of Table~2) can produce a "pseudo-continuum" 
(see e.g. Fig.~2a between 42 and 58 \AA), which bias the determination of the real but very weak continuum.
Several fits to the LETGS spectrum were made: a) to the total spectrum, b) to the total spectrum with selected lines in the
wavelength range from 40 to 100 \AA, to limit the influence of the inaccuracy of atomic data of Ne-, Mg-, and Si- L-shell lines, and
c) to a line spectrum with lines of Table~1 and lines with a statistical significance $\ga$~4$\sigma$ in the wavelength range 
above 40~\AA\ (see Table~5). During our investigations the absolute (relative to H) 
 abundances turned out to be very sensitive to the selected group of elements introduced in the fit procedure. This is especially true
 for the elements Ar and Ca.  
For these reasons no consistent absolute values of the abundances could be obtained. However, abundance ratios
turn out to be much more robust. Therefore the abundance values are 
normalized to oxygen, and are given relative to their solar photospheric values (Anders \& Grevesse (1989)), except for iron. 
For Fe we use log A$_{Fe}$ is 7.51 (see Drake et al. 1995) instead of 7.67 (Anders \& Grevesse 1989). Here 
log A$_{Fe}$ is the logarithmic of the Fe-abundance relative to log A$_H$=12.0. 
The abundances presented in Table~4 are derived assuming the same abundances for the three temperature components. 
These are averaged over the different fits, together with their least-squares-fit standard deviations (within parentheses).

We obtain abundance values between
0.7 and 2.4 relative to Oxygen (e.g., some enhancement for Ne and Si). However, apart from statistical errors these values are also sensitive to systematic errors, due to 
changes in values of the solar photospheric abundances, where uncertainties of a factor of 2 cannot be excluded 
(e.g., Prieto et al. 2001; Grevesse \& Sauval 1998).  
So we cannot obtain indications for a significant FIP effect (enhancement of elements with a low First Ionization Potential)
as found for the solar corona (e.g., Feldman et al. 1992).
This confirms the conclusions by Drake et al. (1995), based on relative abundances from
EUVE observations.
The abundances of C and N, relative to O are somewhat higher than the values obtained in the solar 
photosphere (Anders \& Grevesse 1989). In the EUVE observations by Drake et al. (1995) no suitable C- and N-lines were present 
to constrain (relative) abundances.\\
Values for $n_\mathrm{e}$, given in Table~4, have been obtained by fitting to the O VII and N VI triplet lines. The C~V lines have been omitted from this 
procedure because their intensities are sensitive for the stellar UV-radiative field, mimicing higher densities
 (Ness et al. 2001; Porquet et al. 2001).

\begin{table}
\caption{Best-fit parameters for a 3-$T$ CIE model fit. Elemental abundances 
for the three instruments are given normalized to oxygen and relative to solar photospheric values
 given by Anders \& Grevesse (1989), except for Fe$^{a}$.
1$\sigma$ uncertainties are given in brackets}
\begin {center}
\begin{tabular}{l|ll}
\hline
Parameter   &  LETGS  & RGS+MOS\\
\hline
log $N_\mathrm{H}$ [cm$^{-2}$] & 18.06$^b$ &18.06$^b$\\
$T_1$ [MK] & 0.63(.10) &--- \\
$T_2$ [MK] & 1.21(.07) &1.65(.15) \\
$T_3$ [MK] & 2.26(.12) & 2.68(.22)\\
$EM_1$ [10$^{50}$cm$^{-3}$] & 0.41(.14) & ---\\
$EM_2$ [10$^{50}$cm$^{-3}$] & 2.45(.27) & 3.0(.20)\\
$EM_3$ [10$^{50}$cm$^{-3}$] & 1.72(.29) & 0.9(.18)\\
$n_\mathrm{e}$$_2$[$10^{10}$cm$^{-3}]$      & 1.4 $^{+1.5}_{-0.6}$&1.5$^{+2.0}_{-0.6}$\\
$n_\mathrm{e}$$_3$[$10^{10}$cm$^{-3}]$      & 0.2 $^{+0.8}_{-0.2}$& ---\\
O/H & 0.68(0.38)&0.76(0.33)\\
C/O&1.38(.24)&1.45(.29)\\
N/O&1.33(.10)&1.47(.5)\\
O/O&1.0      &1.0       \\
Ne/O&1.49(.21)& 1.53(.27)\\
Mg/O&1.1(.5)& ---\\
Si/O&1.56(.36)&---\\
S/O&0.69(.15)&---\\
Fe/O&0.97(.31)&1.47(.22)\\
Ni/O&2.39(.27)&---\\
\hline
\end{tabular}
\end{center}
\begin{flushleft}
{
\begin{description}
\item $^a$ In logarithmic units, with log$_{10}$H=12.00;C=8.56;N=8.05;O=8.93;Ne=8.09;Mg=7.58;\\
Si=7.55;S=7.21;Ar=6.56;Ca=6.36;Fe=7.51 (see text);Ni=6.25.
\item $^b$ see Linsky et al. (1995).

\end{description}
}
\end{flushleft}
\end{table}

\subsubsection{Temperature dependent emission measure modeling}
To show the connectivity of the different temperature components we applied
a differential emission measure (DEM) model of Procyon's corona using the various 
inversion techniques offered by SPEX (see Kaastra et al. 1996b). We applied the abundances 
obtained in Sect.~3.2.1. In Fig.~3 we give the results
based on the regularisation method. Other inversion methods (smoothed clean, or 
polynomial) give statistically comparable results. 
The DEM modeling has been applied separately to RGS+MOS and to LETGS. 

\begin{figure}
\center{\psfig{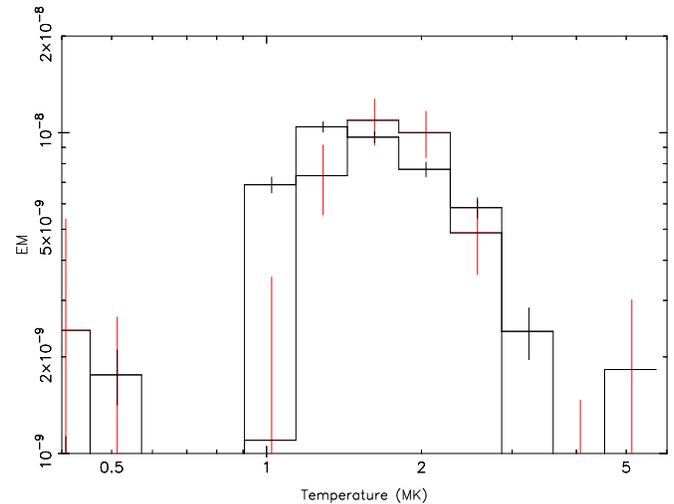}}
\caption[]{EM ($n_\mathrm{e} n_\mathrm{H} V$ in $10^{64} m^{-3}$) for RGS (red) and LETGS (black), using the regularisation
algorithm. The relative abundances given in Table~4 have been applied}

\end{figure}

As a result we find a dominant emission measure of
the order of $ 10^{50}$~cm$^{-3}$ between 1-3~MK. The total emission measures are $3.5(.3) \times 10^{50}$~cm$^{-3}$ for RGS+MOS
 and $4.5(.2) \times 10^{50}$~cm$^{-3}$ for LETGS (in line with the multi-temperature fitting). 
Fig.~3  allows us to conclude that there is no significant amount of $EM$ at $T \ga$ 4~MK in the corona of Procyon. Schmitt et al. (1996) give 
an upper limit of 6~MK, based on EUVE observations.
The $EM$ observed at different times as well as lines fluxes in Table~1 show no significant variability.

Fig.~4 shows fit residuals of parts of the LETGS spectrum fitted using this temperature-dependent emission measure 
modeling, i.e. applying the model of Fig.~3 (LETGS). 
From Fig.~4a we recognize large deviations in residual due to model insufficiencies and a pseudo-continuum 
(most fit residuals positive) due to the lack of weaker lines in current atomic databases in this wavelength range.  
Clear from Fig.~4b are the succeeding large positive and negative residuals around 148 and 171 \AA, due to 
wavelength deviations of lines in the spectrum and the model.

\begin{figure}
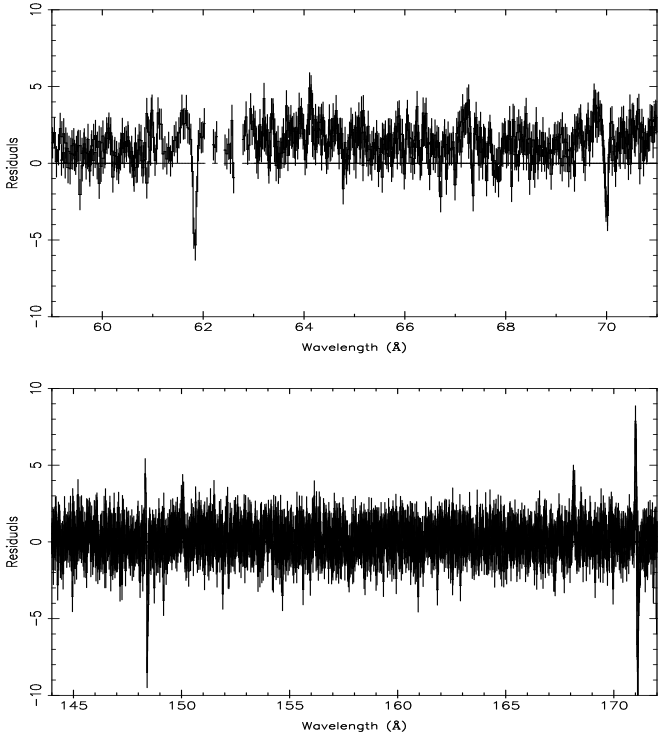

\center{\psfig{figure=H3110F4a.ps,angle=-90,height=5.0truecm,width=8.7truecm,clip=}}
\center{\psfig{figure=H3110F4b.ps,angle=-90,height=5.0truecm,width=8.7truecm,clip=}}
\caption[]{Fit residuals ((observed - model)/error) of parts of the LETGS spectrum}

\end{figure}

\begin{table}
\caption{Observed line fluxes and fluxes obtained from the emissivity from the model} 
\begin {center}
\begin{tabular}{|l@{\ }l@{\ }|l@{\ }c@{\ }c@{\ }r@{\ }|}
\hline 
\multicolumn{2}{|c|}{LETGS }&\multicolumn{4}{c|}{Line identifications}\\
\hline
\multicolumn{2}{|c|}{Observed}&\multicolumn{4}{c|}{Model}\\
\hline
$\lambda$(\AA) &  flux$^a$&$\lambda$(\AA)&3-T flux$^b$&DEM-flux$^b$ & Ion  \\
18.972 & 1.83(15) & 18.973&1.93&1.80&   N VII \\
21.597 & 3.01(25) & 21.602&3.35&2.94&   N VII \\
24.790 & 0.80(14) & 24.781&0.76&0.70&   N VII \\
33.731 & 4.02(32) & 33.736&4.64&4.19&   C VI \\
40.263 & 2.29(36) & 40.270&2.03&1.40&   C V \\
40.718 & 1.88(42) & 40.730&1.33&   &   C V \\
41.475 & 1.07(29) & 41.470&0.68&   &   C V \\ 
       &      & 41.480&0.28&0.31& Ar IX \\ 
43.743 & 0.54(8) & 43.740&0.54&0.94& Si XI \\ 
44.150 & 0.67(10) & 44.165&0.64&0.66& Si XII \\
47.242 & 0.46(8) & 47.280&0.19&0.17&  Mg X \\ 
47.452 & 0.48(8) & 47.500&0.61&0.60&  S IX \\ 
47.642 & 0.49(8) & 47.654&0.20&0.31&   S X  \\
49.207 & 1.44(14) & 49.220&0.43&0.95& Si XI \\ 
       &      & 49.180&0.41&0.37&Ar IX \\
50.520 & 1.68(15) & 50.530&1.48&1.55& Si X \\ 
50.686 & 1.30(14) & 50.690&1.50&1.58&  Si X \\
52.306 & 0.75(11) & 52.300&0.45&0.74& Si XI \\
61.020 & 1.41(25) & 61.050&2.51&1.98& Si VIII \\
61.087 & 1.38(24) &       &    &   & Si VIII$^c$ \\       
63.283 & 0.94(15) & 63.294&1.07&0.93&  Mg X \\
69.646 & 2.03(21) & 69.658&0.85&0.67& Si VIII \\ 
       &      & 69.660&1.14&1.11&Fe XV \\  
74.860 & 1.10(18) & 74.854&0.51&0.57& Mg VIII \\                    
       &      & 74.845&0.26&0.40& Fe XIII \\
75.035 & 1.05(18) & 75.034&0.52&0.57& Mg VIII \\        
77.740 & 1.11(18) & 77.741&0.42&0.37&Mg IX \\
86.765 & 1.13(17) & 86.772&0.71&0.45& Fe XI \\ 
88.087  & 1.68(20)   &88.092&2.33&1.62& Ne VIII \\ 
98.251 & 2.89(34) & 98.260&3.02&2.37& Ne VIII \\
105.20 & 1.22(21) & 105.21&0.32&0.19& Fe IX \\
130.92 & 1.78(39) & 130.94&0.29&0.20& Fe VIII \\
131.21 & 1.78(35) & 131.24&0.41&0.29& Fe VIII \\
148.36 & 11.3(10)  & 148.40&11.0&15.5& Ni XI \\
150.08 & 5.08(60) & 150.10&2.8&2.63&  O VI \\ 
152.11 & 5.60(66) & 152.15&2.9&6.0&Ni XII \\
167.43 & 3.9(10) & 167.49&3.9&2.64 & Fe VIII \\
167.59 & 3.9(12) & 167.66&4.0&2.74& Fe VIII \\
168.13 & 7.4(13) & 168.17&0.3&0.20& Fe VIII \\
168.51 & 5.4(13) & 168.54&2.0&1.42& Fe VIII \\ 
168.90 & 5.0(18) & 168.93&1.3&0.71& Fe VIII \\ 
171.04 & 114(8) & 171.08&100&78& Fe IX \\ 
\hline
\end{tabular}
\end{center}
\begin{flushleft}
{
\begin{description}
\item $^a$ Observed flux in 10$^{-4}$ photons/cm$^2$/s
with in parentheses 1$\sigma$ uncertainty in the last digits.
\item $^b$ Model flux in 10$^{-4}$ photons/cm$^2$/s.
\item $^c$ Sum of two Si VIII lines to be compared with model flux.
\end{description}
}
\end{flushleft}
\end{table}

\subsection{Consistency checks using individual lines}
The question is whether the model insufficiencies influence our conclusions about temperatures,
emission measures, and abundances as obtained in Sect.~3.2. Therefore we have also compared observed and model line fluxes.
The advantage of this individual line approach is that we can select strong and unblended lines,
for which the theoretical emissivities are quite well established.

For the short-wavelength region this is done for all lines (Table~1), while for the longer wavelength range only
lines with a statistical significance $\ga$~4$\sigma$ were used. For the latter the fluxes have been compared with
the 3-T model as well as with the results from the DEM model. The values are given in Table~5. 
This table shows generally a good agreement between the observed flux and the 3-T flux and the flux from the DEM-modeling, summed over 
the T-bins.
Most striking are the deviations for the Fe VIII lines around 131 and 168 \AA. This is definitely due to a large deficiency
in the atomic data used. From atomic physics grounds the line at 168.13 \AA\ is the stronger, as observed in the 
spectrum and in laboratory experiments (Wang et al. 1984), but in our code this line turns out to be the 
weakest.\footnote{\textrm{Recent calculations for Fe~VIII by Griffin et al. (2000) give for our observed Fe~VIII ratio 131/168=0.14
a temperature of $\sim$~2.5~MK, near the high-energy limit of the DEM distribution.}} Another interesting 
feature is the contamination of the forbidden C V line - which is often used for density diagnostics - with Ar IX. Another clear
example of blending is the line at 74.860 \AA\ which contains Mg VIII and Fe XIII.

We have measured line ratios of density-sensitive He-like triplets from the LETGS and RGS spectra, taking into account the
photo-exciting UV flux (Porquet et al. 2001). Our results are consistent in both instruments ($n_\mathrm{e} \approx 10^{10}$~cm$^{-3}$)
and similar to those of Ness et al. (2001) and our values given in Table~4. These results are also comparable to values obtained by 
Schrijver et al. (1995) and Schmitt et al. (1996) and to values for the 
Sun (Drake et al. 2000).\\

\section{Conclusions}
The RGS and LETGS spectra of the corona of Procyon below 40 \AA\ are dominated by the H- and He-like 
transitions of C, N, and O and by Fe XVII lines. 
Above 40 \AA\ the LETGS spectrum shows many L-shell lines of e.g., Ne, Mg, and Si, 
together with lines of Fe VIII-XIII of which the Fe IX line at 171.075 is  very prominent.
All methods applied in Sect.~3.2 to the spectra of the RGS+MOS and the LETGS show temperatures 
of the corona of Procyon between 1--3 MK. 
No indication for a considerably higher temperature component ($T \ga$~4~MK) is found.
The total $EM$ obtained using RGS and LETGS is about $4.1(.5) \times 10^{50}$~cm$^{-3}$. The $EM$ 
distribution shows a smooth continuous structure without
separated peak structures.
Our results improve on those of Schmitt et al. (1996) who obtain an $EM$ distribution with a maximum temperature
around 1.6~MK and a cutoff beyond 6.3~MK.
No significant variability of the coronal conditions took place between the observations by RGS and LETGS.

The abundances of C and N, relative to O are somewhat higher ($\sim$ factor 1.5) than the values obtained in the solar 
photosphere (Anders \& Grevesse 1989). The Fe abundance is about 1--1.5 $\times$ solar.
No significance for a FIP effect, as observed in the solar corona (Feldman et al. 1992), is found.
The same was concluded by Drake et al (1995), based on EUVE observations.
This result is an exception of the trends found by Audard et al. (2001c) for RS CVn systems and by 
G\"udel et al. (2001c) for
solar analogs. These authors have found indications for the evolution from an inverse FIP effect for highly 
active stars - via the absence of a FIP effect in intermediately
active stars - towards a normal FIP effect for less active stars. Clearly, the weakly active star 
Procyon does not fit into this
picture.

\begin{acknowledgements}

The Space Research Organization Netherlands (SRON) is supported
financially by NWO.  
The PSI group acknowledges support from the Swiss National Science Foundation 
(grant 2100-049343).
We are grateful to the calibration teams of the instruments on board
XMM-Newton and Chandra. We thank Nancy Brickhouse and Jeremy Drake for their 
efforts to obtain a long LETGS observation.
Finally, we are grateful to the referee for helpful comments.
\end{acknowledgements}

\end{document}